\documentclass[aps,prl,twocolumn,superscriptaddress]{revtex4-2}
\usepackage[dvips]{graphicx}
\usepackage[usenames]{color}
\usepackage[normalem]{ulem}
\usepackage{soul}

\definecolor{ultramarine}{rgb}{0.07, 0.04, 0.56}

\begin{document}


\title{New spin-polarized electron source based on alkali-antimonide photocathode}


\author{V.S. Rusetsky}
\affiliation{Rzhanov Institute of Semiconductor Physics, Siberian Branch, Russian Academy of Sciences, Novosibirsk, 630090, Russia.}
\affiliation{CJSC "Ekran FEP", Novosibirsk, 630060, Russia.}

\author{V.A. Golyashov}
\affiliation{Rzhanov Institute of Semiconductor Physics, Siberian Branch, Russian Academy of Sciences, Novosibirsk, 630090, Russia.}
\affiliation{Synchrotron radiation facility SKIF, Boreskov Institute of Catalysis, Siberian Branch, Russian Academy of Sciences, Kol'tsovo, 630559, Russia.}
\affiliation{Novosibirsk State University, Novosibirsk, 630090 Russia.}

\author{S.V. Eremeev}
\affiliation{Institute of Strength Physics and Materials Science, Tomsk, 634055, Russia.}

\author{D.A. Kustov}
\affiliation{Rzhanov Institute of Semiconductor Physics, Siberian Branch, Russian Academy of Sciences, Novosibirsk, 630090, Russia.}

\author{I.P. Rusinov}
\affiliation{Tomsk State University, Tomsk, 634050, Russia.}

\author{T.S. Shamirzaev}
\affiliation{Rzhanov Institute of Semiconductor Physics, Siberian Branch, Russian Academy of Sciences, Novosibirsk, 630090, Russia.}
\affiliation{Novosibirsk State University, Novosibirsk, 630090 Russia.}

\author{A.V. Mironov}
\affiliation{CJSC "Ekran FEP", Novosibirsk, 630060, Russia.}

\author{A.Yu. Demin}
\affiliation{CJSC "Ekran FEP", Novosibirsk, 630060, Russia.}

\author{O.E. Tereshchenko}
\email[]{teresh@isp.nsc.ru}
\affiliation{Rzhanov Institute of Semiconductor Physics, Siberian Branch, Russian Academy of Sciences, Novosibirsk, 630090, Russia.}
\affiliation{Synchrotron radiation facility SKIF, Boreskov Institute of Catalysis, Siberian Branch, Russian Academy of Sciences, Kol'tsovo, 630559, Russia.}
\affiliation{Novosibirsk State University, Novosibirsk, 630090 Russia.}

\date{\today}

\begin{abstract}
New spin-dependent photoemission properties of alkali antimonide semiconductor cathodes are predicted based on the detected optical spin orientation effect and DFT band structure calculations. Using these results, the Na$_2$KSb/Cs$_3$Sb heterostructure is designed as a spin-polarized electron source in combination with the Al$_{0.11}$Ga$_{0.89}$As target as a spin-detector with spatial resolution. In the Na$_2$KSb/Cs$_3$Sb photocathode, spin-dependent photoemission properties were established through detection of high degree of photoluminescence polarization and high polarization of the photoemitted electrons. It was found that the multi-alkali photocathode can provide electron beams with emittance very close to the limits imposed by the electron thermal energy. The vacuum tablet-type sources of spin-polarized electrons have been proposed for accelerators, that can exclude the construction of the photocathode growth chambers for photoinjectors.

\end{abstract}


\maketitle

In recent decades, the physics of spin-polarized electrons in semiconductors has proven to be not only a fascinating area of basic research, but also a fruitful field of device applications that have had a significant impact on several areas of modern physics. Spin-polarized electron beams with high currents are required by modern and future nuclear physics facilities such as the Electron Ion Collider (EIC), the International Linear Collider \cite{Wang_ERL2017,Aprahamian2015,Michizono2019,Musumeci2018,Pellegrini2012} and for developing new generation of electron-positron collider "Super charm-tau factory" \cite{Piminov2018} with a variety of uses. 

Long lifetime polarized electron sources are also of interest for electron microscopy methods exploiting spin polarization to probe magnetization in materials and nanostructures \cite{Suzuki2010,liang2018}. Probing the spin-resolved electron states of solids, surfaces and nanostructures give direct access to phenomena like magnetism \cite{Vollmer2003}, proximity effects~\cite{Buzdin_RevModPhys2005}, spin–orbit interaction \cite{Bentmann2021} and related spin texture in low-dimensional systems \cite{Maas2016}, which are regarded as potential functional systems for developing low-energy fast spintronic sensors or logic devices.

Almost all modern electron sources for highly spin-polarized electron beams in accelerator physics and electron microscopy rely on photocathode materials based on III-V (GaAs) technology. The first realization of a GaAs spin-polarized source was made by Pierce et. al. \cite{PIERCE1975} in 1975 based on both the phenomenon of optical orientation of electron spins in semiconductors (creation of spin-oriented carriers with absorption of circularly polarized light) and the discovery of negative electron affinity (NEA) activation of $p$-type semiconductor surfaces. The appearance of charge carriers with oriented spins upon absorption of a circularly polarized light was originally studied by Garvin et al. \cite{Garvin1974} and by Lampel and Weibush \cite{LAMPEL1975}. Despite the widespread use of GaAs as a photocathode, a serious disadvantage of the GaAs cathode is its sensitivity to residual vacuum gases and, as a consequence, a short lifetime (or very low quantum efficiency) \cite{Friederich2019}. A more stable semiconductor electron emitter sources are based on the alkali metal antimonide cathodes \cite{Musumeci2018}. 

Alkali antimonide semiconducting materials are of great interest for  single photon detection in photomultiplier devices \cite{Bergevin2015,Wright2017,Lyashenko2020} and as electron sources for the generation of high brightness electron beams for the next generation light sources like Energy Recovery Linacs (ERL's) and Free Electron Lasers ~\cite{DOWELL2010,Michizono2019,Scholz2018}. The advantage of these photocathodes is provided by their faster temporal response compared to III-V photoemitters.  
It was also demonstrated that the multi-alkali photocathode can provide bright electron beams with intrinsic emittance very close to the limits imposed by the electron thermal energy (0.22 mm-mrad/(mm rms) corresponding to a Mean Transverse Energy (MTE) of 30 meV at room temperature) \cite{Cultrera2016}. 
 The epitaxial growth of thin films of the high-efficiency photocathode  alkali antimonide materials by molecular-beam epitaxy opens the way to a significant increase in brightness and efficiency near the threshold by reducing surface disorder \cite{Yamaguchi2017,Feng2017,Parzyck_PRL22-Cs3Sb}.
 All this leads to the fact that some groups switch from GaAs photocathode to alkaline antimony photocathode for DC gun or RF gun \cite{Wang_ERL2017}.

First attempts to use photosensitive materials, such as Cs$_3$Sb, K$_2$CsSb, Na$_2$KSb, and Na$_2$KSb:Cs for electron sources have been performed since the 1930s \cite{Gorlich1936, Sommer1955, Spicer1958}, however, surprisingly, the possibility of their use as a source of spin-polarized electrons has not been studied so far.

This Letter presents the first observation of spin-polarized photoemission from alkali-antimonide semiconductor heterostructures.
The vacuum spin-photodiode consisting of semiconductor Na$_2$KSb photocathode and Al$_{0.11}$Ga$_{0.89}$As spin-detector, both activated to NEA states, is implemented and high spin polarization of photoemitted electrons from Na$_2$KSb measured in spectral and image modes is demonstrated.


From a technological point of view,  knowledge of the electronic structure and spin texture is essential in order to understand the spin-polarized photoemission properties. 
\begin{figure}
\includegraphics[width=\columnwidth]{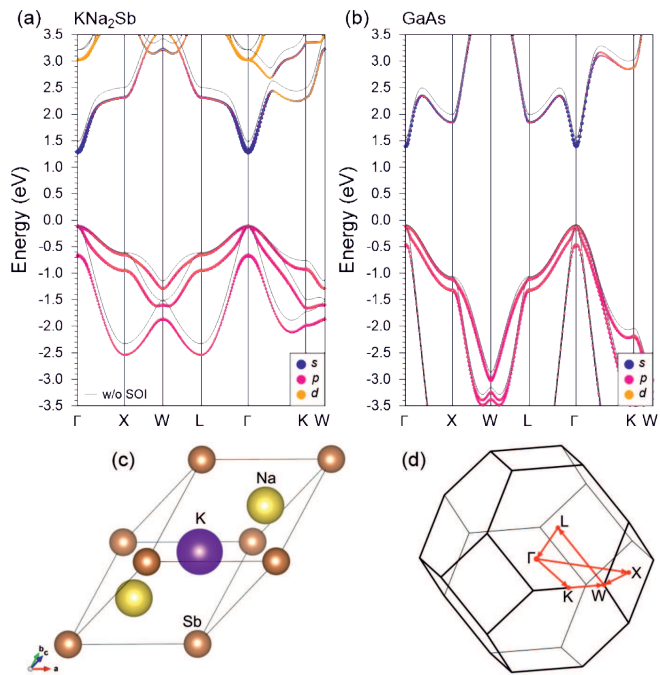}
\caption{Calculated bulk electronic band structures of (a) Na$_2$KSb and (b) GaAs. Gray lines show spectra calculated without SOI included. Colored dots show orbital character of the states (see keys). (c) Balls-and-sticks representation of the primitive cell of cubic Na$_2$KSb  and (d) corresponding bulk Brillouin zone with the high-symmetry points and the path connecting them, along which the spectra are shown in panels (a) and (b). }
\label{band_structure}
\end{figure}
The calculated dispersion of the electron bands of cubic Na$_2$KSb  is shown in Fig.~\ref{band_structure}(a). In the density functional theory (DFT) calculations performed with the Vienna Ab Initio Simulation
Package (VASP) \cite{VASP1,VASP2}, where the generalized
gradient approximation (GGA-PBE) \cite{perdew} for the exchange-correlation potential was applied and the interaction between the ion cores
and valence electrons was described by the projector augmented-wave method~\cite{PAW1,PAW2}, the alkali metals $p$ orbitals were treated as valence electrons. To obtain realistic bulk gap the Slater-type DFT-1/2 self-energy correction method \cite{DFT12_1,DFT12_2} with a partially (quaternary) ionized antimony potential was applied.

At the first glance, one can see a very similar band structure for Na$_2$KSb and GaAs (calculated within similar approach, Fig.~\ref{band_structure}(b)). Gray lines in Fig.~\ref{band_structure} show the spectra calculated with switched-off spin-orbit interaction (SOI) for both Na$_2$KSb and GaAs, where three topmost valence bands are degenerate at $\Gamma$ and the band gaps are 1.58 and 1.62 eV, respectively. The calculated energy band dispersion of the non-relativistic electronic structure of Na$_2$KSb and symmetries of the bands in the center of the Brillouin zone agree well with the earlier calculations performed within the localized spherical wave method~\cite{Ettema_PRB2000} with the exception of smaller band gap in the latter case (0.9 eV), as well as with results obtained within $GW$ approximation (1.51 eV)~\cite{Amador2021} with which there is a fine agreement. Switching SOI on leads to a slight narrowing of the band gaps, to 1.41 and 1.51 eV, respectively. However, the main changes occur in the valence band, that is by 90\% is determined by the $p$ states of Sb(As). The heavy-hole and light-hole valence bands remain degenerate while the third band splits off. This splitting is about twice larger in Na$_2$KSb (0.55 eV vs. 0.34 eV in GaAs). These bands form initial states in the photoabsorption process. 

Thus, the similarity of the band structures of the Na$_2$KSb and GaAs allows us to conclude that the optical orientation of electron spins is possible in the Na$_2$KSb, and, consequently, the photoemission of spin-polarized electrons.


\begin{figure*}
\includegraphics[width=\textwidth]{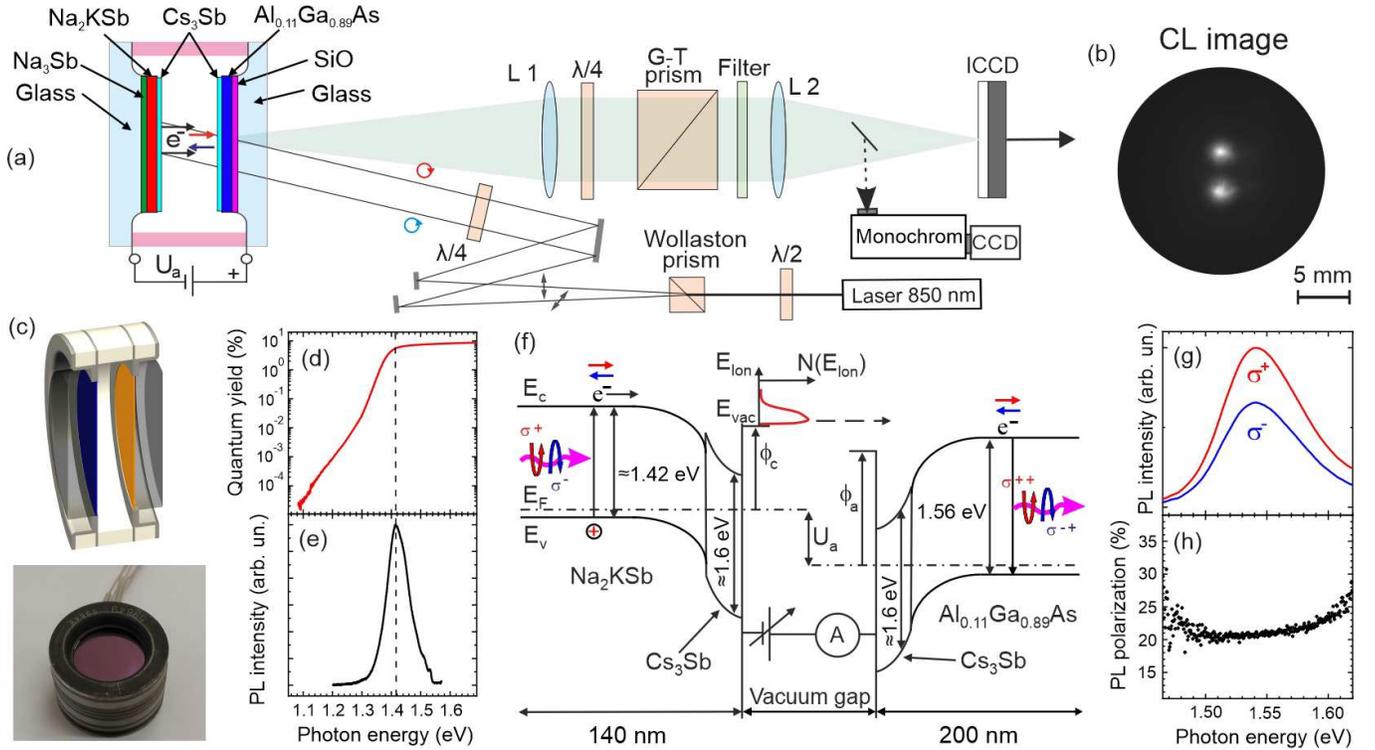}
\caption{Experimental setup. (a) Schematic drawing of optical setup for spatial CL polarization detection with a compact vacuum photodiode for the spin polarization measurements. Electrons emitted by $\hbar\nu$-photon excitation from the Na$_2$KSb source are injected into an Al$_{0.11}$Ga$_{0.89}$As target.
(b) Image of CL intensity distribution in the anode Al$_{0.11}$Ga$_{0.89}$As structure under injection of 2 eV electrons emitted from the photocathode at room temperature. (c) Cross-sectional view and a photograph of the assembled photodiode from the Na$_2$KSb photocathode side, which can be used as a tablet-type source of spin-polarized electrons for accelerators.
(d) Quantum yield as a function of the incident photon energy and (e) photoluminescence spectra from Na$_2$KSb photocathode excited with the 1.59 eV (780 nm) laser diode. (f) Energy band diagram of both the Na$_2$KSb source and the Al$_{0.11}$Ga$_{0.89}$As target with the NEA separated by a vacuum gap. $E_c$ is the conduction band, $E_v$ is the valence band and $E_{\rm F}$ is the Fermi level. (g) Circularly polarized ($\sigma^+$, $\sigma^-$) components of the PL spectra of Al$_{0.11}$Ga$_{0.89}$As target excited with the circular polarized light with 1.67 eV (740 nm) photon energy. (h) The corresponding circular polarization degree of the PL emission determined as $P_{\rm PL}=(I_{\sigma^+}-I_{\sigma^-})/(I_{\sigma^+}+I_{\sigma^-})$}.
\label{experiment}
\end{figure*}

In our previous work \cite{Tereshchenko2021} we developed an image-type spin detector prototype for measuring the normal (to the surface of the detector) component of the electron beam polarization. This detector is based on the injection of spin-polarized free electrons from GaAs/(CsO) cathode into a heterostructure with GaAs/Al$_{x}$Ga$_{1-x}$As quantum wells (QWs) and recording the circularly polarized cathodoluminescence (CL) with spatial resolution. A flat vacuum photodiode composed of two effective NEA semiconductor electrodes was designed and studied in Refs.~\cite{Rodionov2017,Tereshchenko2017,GOLYASHOV2020,Rusetsky2021}. In this work we demonstrate that alkali antimonide photocathode  can be utilized as an effective source of spin-polarized electrons and the anode - Al$_{0.11}$Ga$_{0.89}$As/Cs$_3$Sb heterostructure, as the spin detector. Schematic presentation of the compact vacuum photodiode with a photo of the device and investigation of spin-dependent injection are shown in Fig.~\ref{experiment}(a). 
The photocathode consists of an active 140 nm Na$_2$KSb layer grown on the glass of the input window and activated to NEA by Cs$_3$Sb layer. The semiconductor anode (target) comprises 200 nm Al$_{0.11}$Ga$_{0.89}$As layer with SiO coating, through which the structure is bonded to the glass of the output window. The final step of the cleaning procedure for the anode was carried out inside a glove box flooded with pure nitrogen, in which the anode was chemically treated in a solution of HCl in isopropanol \cite{Tereshchenko1999}. The cleaned Al$_{0.11}$Ga$_{0.89}$As anode surface was also activated to the NEA state by coadsorption of cesium and antimony. Similar activation layer was utilized earlier for GaAs~\cite{Bazarov2020}. The photocathode and anode were plane-parallel mounted in an air-tight manner on the opposite flat sides of a cylindrical alumina ceramics body. The diameters of the cathode and anode were 18 mm with a 1.0 mm gap between the electrodes.
To compare spin-dependent photoemission properties of Na$_2$KSb/Cs$_3$Sb photocathode with the GaAs one,  we also constructed and studied the vacuum spin-photodiode with GaAs/(Cs,O) cathode and the same Al$_{0.11}$Ga$_{0.89}$As/(Cs,O) anode. Based on the developed vacuum diode (Fig.~\ref{experiment}(c)), the tablet-type sources of spin-polarized electrons can be manufactured and used for accelerators. The evacuated photocathode source can be opened in the photoinjector, thus potentially eliminating the construction of photocathode growth chambers (see Supplemental Material, Fig.~S1 \cite{SM}).

\begin{figure*}
\includegraphics[width=\textwidth]{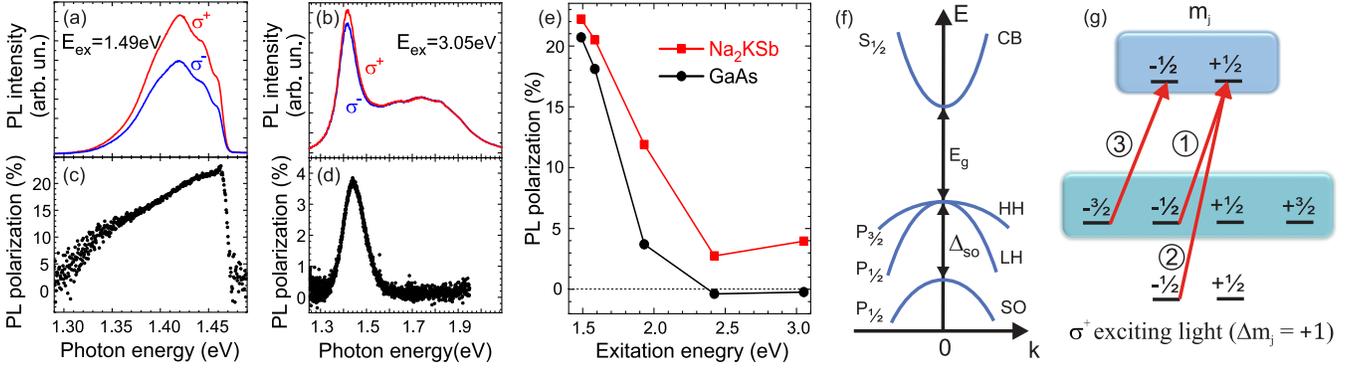}
\caption{Circularly polarized ($\sigma^+$, $\sigma^-$) components of the PL spectra of Na$_2$KSb photocathode excited with the circular polarized emission of laser diode with (a) 1.49 eV (780 nm) and (b) 3.05 eV (405 nm) excitations. The peaks at 1.42 eV in panels (a) and (b) correspond to the Na$_2$KSb band gap transition. (c) and (d) The corresponding circular polarization degree of the PL emission determined as  $P_{\rm PL} =(I_{\sigma^+} - I_{\sigma^-})/(I_{\sigma^+} + I_{\sigma^-})$. (e) The circular polarization degree of the PL emission of Na$_2$KSb compared with GaAs (0.5 $\mu$m thickness) as a function of photon excitation energy. 
(f) Schematic band structure of Na$_2$KSb near the $\Gamma$-point illustrating the band gap (E$_g$), the spin-orbit splitting ($\Delta_{\rm SO}$), the conduction band (CB), the heavy hole (HH) band, the light hole (LH) band and the spin-orbit split-off (SO) band. 
(g) Optical selection rules for transitions between the valence band and the conduction band for right circularly polarized light ($\sigma^+$). The numbers in circles denote the relative transition probabilities.}
\label{PL} 
\end{figure*}

Quantum yield (QY) of Na$_2$KSb photocathode measured in the transmission mode as a function of the incident photon energy in combination with photoluminescence spectrum are shown in Fig.~\ref{experiment}(d). The QY is calculated as the ratio of the registered electrons per incident photons. The quantum yield has the threshold at about 1.4 eV and increases slightly in the range 1.4-2.5 eV reaching QY maximum at about 15~\%. 
The photoluminescence (PL) spectrum of the Na$_2$KSb photocathode recorded at a room temperature (Fig.~\ref{experiment}(e)) shows a peak with the energy of 1.42 eV (887 nm) which corresponds to the Na$_2$KSb  optical band gap transition in agreement with DFT calculation (Fig.~\ref{band_structure}~(a)). One can see that PL peak position coincides well with the photoemission threshold of QY spectra, meaning that the red edge in photoemission spectrum is determined by the energy gap rather than the work function of Na$_2$KSb/Cs$_3$Sb. Thus, the deposition of Cs$_3$Sb layer on Na$_2$KSb leads to formation of the NEA state as follows from QY, PL (Fig.~\ref{experiment}\~(d, e)) and energy distribution curves (EDC) measurements (Fig.~\ref{EDC}).  
These results, together with earlier data on Cs$_3$Sb photoemission properties \cite{Spicer1958}, allow us to plot the band diagram shown in (Fig.~\ref{experiment}(f)).

To characterize the Al$_{0.11}$Ga$_{0.89}$As/Cs$_3$Sb target as a spin-detector first we  measured the circularly polarized PL and determined the spectral dependence of the degree of circular polarization of PL (Fig.~\ref{experiment}(g)). For $p$-Al$_{0.11}$Ga$_{0.89}$As layer of 200 nm thickness under excitation with photon energy $\hbar\omega$ = 1.67 eV the degree of PL polarization is about 22~\%  that is close to the 
potentially feasible value of 25\%. The results of polarized PL measurements show  that $p$-Al$_{0.11}$Ga$_{0.89}$As can be used for spin polarimetry applications based on the optical detection of the free-electron spin polarization similar to QWs structure used in Ref.~\cite{GOLYASHOV2020}.

\begin{figure*}
	\includegraphics{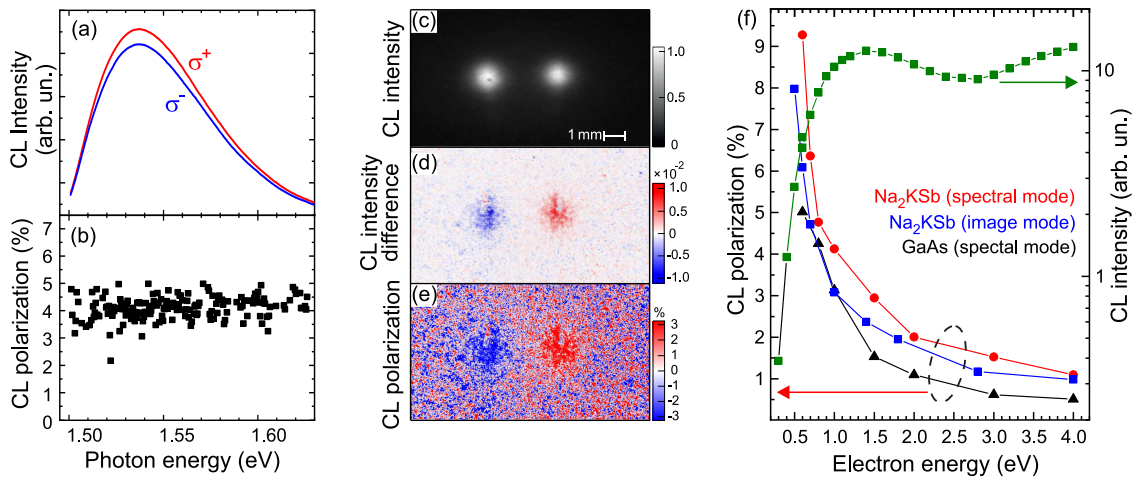}
	\caption{(a) Circularly polarized ($\sigma^+$, $\sigma^-$) components of the CL spectra excited by the injection of spin polarized electrons emitted from the Na$_2$KSb/Cs$_3$Sb cathode at the accelerating voltage of 1.0 V. (b) The circular polarization degree of the CL emission determined as  $P_{\rm CL} =(I_z^\uparrow - I_z^\downarrow)/(I_z^\uparrow + I_z^\downarrow)$. (c) Image of the spin-integrated CL intensity, (d) Difference between ‘spin-up’ and ‘spin-down’ CL intensities ($I_z^\uparrow - I_z^\downarrow$), and (e) 2D distribution of the derived CL circular polarization degree $P_z$. The images were taken at an accelerating voltage of 1 V and $T$ = 300 K. 	 
	(f) Left axis: dependence of the CL circular polarization degree on the energy of injected spin-polarized electrons for the Na$_2$KSb photocathode compared with the GaAs/(Cs,O) photocathode where a similar Al$_{0.11}$Ga$_{0.89}$As/(Cs,O) spin-detector is used. Right axis: dependence of the integrated nonpolarized CL (logarithmic scale) on
the energy of injected electrons (green squares).  }
\label{CL} 
\end{figure*}


In order to test the polarization properties of the Na$_2$KSb/Cs$_3$Sb photocathode, we measured circularly polarized spectra ($\sigma^+$, $\sigma^-$) of photoluminescence in the optical orientation mode. The polarized PL excited with the circular polarized emission of laser diodes with photon energy $\hbar\omega$ = 1.49 eV (850 nm) and $\hbar\omega$ = 3.05 eV (405 nm) is shown in Fig.~\ref{PL} (a) and (b), respectively. The corresponding spectral dependence of the circular polarization degree for the PL emission is shown in the bottom panels in Fig.~\ref{PL} (c,d). As can be seen, even at excitation energy of 3.05 eV the PL polarization is not zero (4~\%). Note that the broad unpolarized peak centered at 1.8 eV in Fig.~\ref{PL} (b) corresponds to the glass PL emission \cite{Spin2021}. Dependence of the PL circular polarization degree on the excitation photon energy in Na$_2$KSb in comparison with GaAs is shown in Fig.~\ref{PL}~(e). At photon energy $\hbar\omega$ = 1.49 eV the PL polarization showing a maximum, 23~\%, decreases rapidly with rising in the excitation energy up to $\sim$2.4 eV. However, in contrast to GaAs, where PL polarization becomes zero, in Na$_2$KSb it remains about 4~\% at higher photon excitation energies. 
Similarity in the polarization dependence of the PL at low excitation energies and in the band structures of Na$_2$KSb and GaAs (Fig.~\ref{band_structure} (a,b)) suggests that the optical selection rules for these two semiconductors are the same. 
Schematic band structure of Na$_2$KSb in the vicinity of the BZ center as derived from the DFT spectrum (Fig.~\ref{band_structure}(a)) is shown in Fig.~\ref{PL}(f). The conduction band is two-fold degenerate and consists of $s$-like states (total angular momentum $j = 1/2$). The valence bands, on the other hand, consist of $p$-like states with a two-fold degenerate heavy-hole and light-hole subband ($j = 3/2$), and a two-fold degenerate split-off (SO) band ($j = 1/2$) separated by the spin-orbit energy $\Delta_{\rm SO} \cong 0.55$ eV in accordance with DFT results. The Fig.~\ref{PL}~(g) shows the optical transitions in Na$_2$KSb between the valence band and the conduction band. The circled numbers indicate the corresponding transition probabilities for right circularly polarized light ($\sigma^+$). For photons with the energy between $E_g$ and $E_g + \Delta_{\rm SO}$ only transitions from the heavy-hole and light-hole states into the conduction band are allowed. These transitions obey selection rules that preserve the angular momentum of the absorbed photon of +1 (or -1 for left polarized light). 
Thus, circularly polarized light can be used to create an unequal occupation of electrons in Na$_2$KSb with the spin orientation parallel and anti-parallel to the propagation direction of the incident light. If the electrons  escape from the multi-alkali photocathode Na$_2$KSb/Cs$_3$Sb into the vacuum, as shown in the band diagram in Fig.~\ref{experiment} (f), we can expect photoemission of spin-polarized electrons.


In order to test the Na$_2$KSb/Cs$_3$Sb photocathode as a source of spin-polarized electrons, the injection of photoemitted electrons into the Al$_{0.11}$Ga$_{0.89}$As/Cs$_3$Sb heterostructure was studied using spin-polarized cathodoluminescence. Earlier it was shown that semiconductor heterostructures based on III-V compounds can be used as spin detectors of free electrons with spectral and spatial resolution \cite{Li2014,Tereshchenko2011,Tereshchenko2015,GOLYASHOV2020,Tereshchenko2021}. The polarized CL spectra measured  under injection of spin-polarized electrons with the energy of 1.0 eV are shown in Fig.~\ref{CL}~(a). The maximum CL intensity corresponds to the radiation emission energy of 1.53 eV (810 nm), coinciding with the PL maximum of Al$_{0.11}$Ga$_{0.89}$As (Fig.~\ref{experiment}~(g)). The degree of spectral circular polarization of the CL emission excited by the electrons with the energy of 1.0 eV is equal to 4.5 \% (Fig.~\ref{CL}(b)).

Recently we demonstrated that semiconductor spin-detector allows to measure polarization with spatial resolution in direct image mode \cite{Tereshchenko2021}. Fig.~\ref{CL}~(c) shows the image of spin-integrated CL intensity from two spots induced by two opposite spin-polarized electron beams injected into the Al$_{0.11}$Ga$_{0.89}$As/Cs$_3$Sb target (Fig.~\ref{experiment}~(a,b)). The image is taken at an accelerating voltage of 1 V and reflects the electron distribution in the cross section. In this case, the CL polarization in the cross section of two electron beams should have the opposite sign and be determined by the polarization of photoemitted electrons exited by absorbed circularly polarized light in the photocathode. Fig.~\ref{CL}~(d) shows the CL intensity difference ($I_z^\uparrow - I_z^\downarrow$). Due to the injection of opposite polarized electron beams and its recombination in Al$_{0.11}$Ga$_{0.89}$As, the emitted CL contains both polarized emission components ($\sigma^+$, $\sigma^-$) with helicity oppositely distributed intensity in the two spots (Fig.~\ref{CL}~(d)): spin-up $I_z^\uparrow$ and spin-down $I_z^\downarrow$. The CL polarization asymmetry image $P_{\rm CL} =(I_z^\uparrow - I_z^\downarrow)/(I_z^\uparrow + I_z^\downarrow)$ is shown in Fig.~\ref{CL}(e). The detected $P_{\rm CL}$ at electron injection with energy of 1 eV  is 3 \% (Fig.~\ref{CL}~(e)).
The dependences of the CL circular polarization degree determined from spectral and spatial distribution on the injected electron energy in the range of 0.6-4 eV are shown in Fig.~\ref{CL}(f).  $P_{\rm CL}$ is maximal, of 9.5 \%, at low kinetic energies (0.6 eV) and monotonically decreases down to $\approx1$ \% as the electron energy increases to 4 eV. Comparing the CL polarization $P_{\rm CL}$ for Na$_2$KSb/Cs$_3$Sb to GaAs photocathode (Fig.~\ref{CL}(f)) one can conclude that the latter is twice less. Assuming that the polarization of photoelectrons from GaAs cathode is in the range of 20-25 \% \cite{GOLYASHOV2020}, we can conclude that the polarization of photoemitted electrons from Na$_2$KSb/Cs$_3$Sb is in the range of 40-50 \%.


Photoinjectors equipped with low intrinsic emittance photocathodes are among the brightest electron sources in use today \cite{Bazarov2009,Musumeci2018}.
One of the limitations of the peak brightness at the source is the photocathode’s transverse momentum spread. 
To study the electron emission process via the electron energy distribution, the photoelectron spectra are measured by differentiating the delay curves using the lock-in technique. This allows measuring photoelectron longitudinal (along the beam) energy distributions $N$($E_{\rm lon}$) by means of the photodiode as a retarding field electron spectrometer \cite{Terekhov1994, Rodionov2017}.
The electron energy distribution curves (EDCs) measured at room temperature for the Na$_2$KSb/Cs$_3$Sb  transmission-mode with 1.38 and 1.26 eV photon energies have the widths of 101 and 35 meV, respectively (Fig.~\ref{EDC}). 
The EDC width for photon excitation of 1.38 eV 
reflects the presence of NEA on the surface with a value of about 100 meV (Fig.~\ref{experiment}~(f)). 
Rapid drop of EDC width to 35 meV with decreasing photon energy proves that the multi-alkali photocathode can provide spin-polarized electron beams with emittance very close to the limits imposed by the electron thermal energy (see Supplemental Material, Fig.~S2 \cite{SM}). Moreover, the QY at 1.26 eV photon energy is equal to 0.03~\%  (Fig.~\ref{experiment}~(d)) that is much higher than in metallic photocatodes \cite{Karkare2020}.

\begin{figure}
	\includegraphics{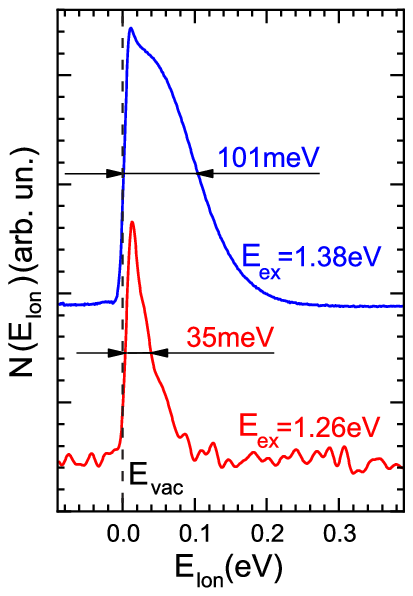}
	\caption{Energy distribution curves measured at photon excitation energies of 1.38 and 1.26 eV.}
		\label{EDC} 
\end{figure}


To summarize, the Na$_2$KSb band structure was found to be very similar to GaAs with the same optical transition rules. However, the SOI gap in the VB of Na$_2$KSb is almost twice larger than that in GaAs. Spin-polarized electrons in the Na$_2$KSb photocathode were obtained by optical pumping with circularly polarized light and showed 23 \% polarization of photoluminescence and 40-50 \% spin polarization of the emitted electrons. The measured energy spreads of electron beams generated with light having energy lower than 1.3 eV (wavelength longer than 900 nm) are approaching the limit imposed by the thermal energy of electrons at room temperature with quantum efficiencies  higher than metallic photocathodes.
    The Al$_{x}$Ga$_{1-x}$As structure has also been shown to be promising as a spatially resolved spin detector for free electrons.  High quantum efficiency and low emittance of spin-polarized electron beams provided by multi-alkali photocathodes make them of great interest for next generation high brightness photoinjectors. Finally, based on the created vacuum photodiode, the tablet-type sources of spin-polarized electrons can be manufactured and used for accelerators without construction of the photocathode growth chambers for photoinjectors.


\section{Supplemental Material}

\subsection{The tablet-type source of spin-polarized electrons }

\begin{figure}
  \includegraphics[width=\columnwidth]{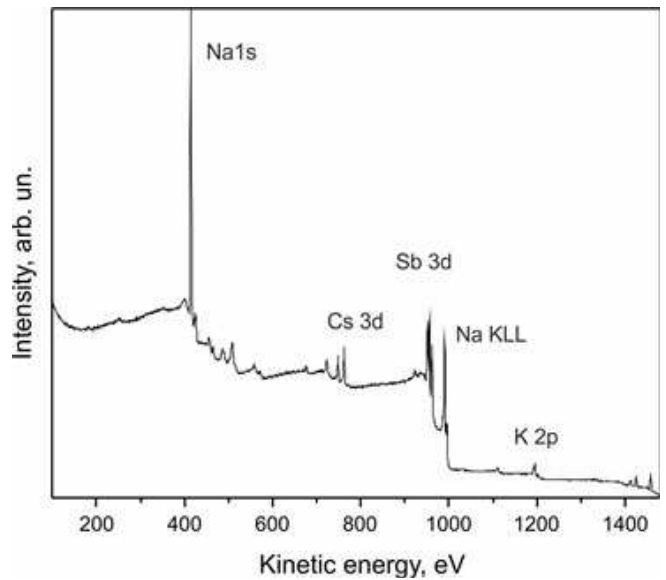}
\caption{XPS spectrum of the Na$_2$KSb/Cs$_3$Sb photocathode measured after opening the photodiode in the preparation chamber of an electron spectrometer.}
 \label{figS1}
\end{figure}

The multialkali photocathodes, owing to their alkali metal content, are extremely sensitive to vacuum contamination and cannot be characterized ex situ. To measure the composition of the cathode surface, a vacuum-sealed diode was opened in an analytical XPS setup. Fig.~\ref{figS1} shows the X-ray photoemission spectrum of the multialkali Na$_2$KSb/Cs$_3$Sb photocathode source after opening the vacuumed photodiode under vacuum condition in the preparation chamber of electron spectrometer. The ratio of alkali components Na:K:Cs and Sb confirms the formation of Na$_2$KSb/Cs$_3$Sb interface. 

Based on the vacuum opening technology, the tablet-type sources of spin-polarized electrons can be manufactured separately and used for accelerators.  

\subsection{Mean transverse energy and emittance}

\begin{figure}
  \includegraphics[width=\columnwidth]{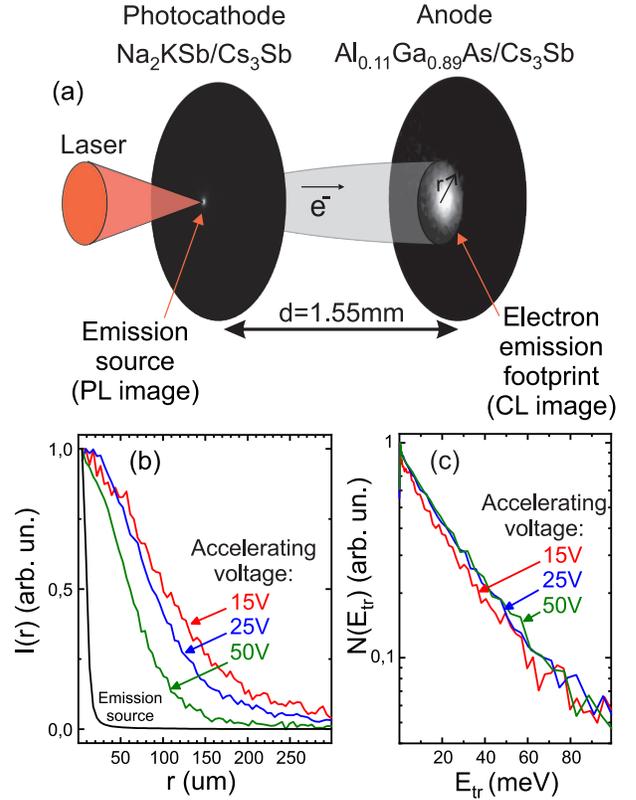}
\caption{(a) Schematic drawing of the experiment. Size of the photoluminescence image exited by the laser light reflects the emission source size. Electrons emitted by laser excitation from the Na$_2$KSb/Cs$_3$Sb photocathode are injected into an Al$_{0.11}$Ga$_{0.89}$As target and generate electron emission footprint, which reflects transverse energy distribution of the emitted electrons. (b) Radial profiles calculated from the recorded emission source image and electron emission footprint . (c) Transverse energy distributions calculated from the radial profiles.}
 \label{figS2}
\end{figure}

To estimate the mean transverse energy (MTE) of the electrons emitted from the multialkali photocathode the transverse energy distribution curves (TEDC) were studied. For the experiment we used the vacuum photodiode described in polarized cathodoluminescence study.
The photocathode was illuminated with 1.26 eV tightly focused laser beam, and its electron emission footprint (cathodoluminescence (CL) image from the anode) was recorded with CCD camera (Fig.~\ref{figS2} (a)). The emission source size was received from the photoluminescence (PL) image recorded from the photocathode. The images of electron emission footprint and emission source were converted to radial profiles $I(r)$ for further calculations (Fig.~\ref{figS2} (b)). 
Since the electron emission footprint is 10 times larger than emission source (Fig.~\ref{figS2} (a, b)) ($\approx$ 200 $\mu$m vs 20 $\mu$m) we consider that the photoelectrons are emitted from a point source. Knowing the distance travelled by the photoelectrons (vacuum gap width) and the voltage through which they have been accelerated (defining their time-of-flight), the transverse energy $E_{tr}$ required to generate the observed emission footprint can be determined:
\begin{equation}
E_{tr}=\frac{1}{2}m\frac{r^2}{t^2},
\label{eq1}
\end{equation}
where $t=\sqrt{2d/a}$ , $a = eU_{acc}/m$ – acceleration, $d$ – vacuum gap width (1.55 mm), $U_{acc}$ – acceleration voltage, $e$ – electron charge, $m$ – electron mass, $r$ – distance from the center of the electron emission footprint. Since $eU_{acc} \gg E_{lon}$, longitudinal energy component is neglected. Radial profiles of electrons $I(r)$ were transformed to TEDC $N(E_{tr})$ (Fig.~\ref{figS2} (c)) using the equation:

\begin{equation}
N(E_{tr})=2\pi r I(r)\left|{\frac{dr}{dE_{tr}}}\right|.
\label{eq2}
\end{equation}

From the fact that TEDC recorded at different acceleration voltages (15, 25, 50 V) are similar one can conclude that electric field is uniform and there is no space charge in the vacuum gap of the photodiode during the experiment. The next step was to calculate the mean transverse energy (MTE) of the electrons emitted from the photocathode. By definition, MTE is given by:

\begin{equation}
MTE=\frac{\int_{0}^{\infty} E_{tr}N(E_{tr}) \,dE_{tr}}{\int_{0}^{\infty} N(E_{tr}) \,dE_{tr}}.
\label{eq3}
\end{equation}

The MTE obtained from the TEDC turned out to be close to the limits imposed by the electron thermal energy (MTE $\approx$ 35 meV). Another parameter used for electron source characterization is the intrinsic emittance $\epsilon_{in}$. It can be calculated from the MTE as:
\begin{equation}
\epsilon_{in}=\sqrt{\frac{MTE}{mc^2}},
\label{eq4}
\end{equation}
where $m$ – electron mass, $c$ – speed of light. The intrinsic emittance corresponding to the MTE measured equals to 0.26 mm$\cdot$mrad/mm. The values of MTE and $\epsilon_{in}$ are close to the results obtained in other works on multialkali photocathodes \cite{Maxson2015,Cultrera2016}.


%

\end{document}